# Nanoscale Thermotropic Phase Transitions Enhance Photothermal Microscopy Signals

A. Nicholas G. Parra-Vasquez[1,2], Laura Oudjedi[1,2], Laurent Cognet[1,2], and Brahim Lounis[1,2,*]

[1] Univ Bordeaux, LP2N, F-33405 Talence, France

[2] CNRS & Institut d'Optique, LP2N, F-33405 Talence, France

ABSTRACT. The photothermal heterodyne imaging technique enabled studies of individual weakly absorbing nano-objects in various environments. It uses a photo-induced change in the refractive index of the environment. Taking advantage of the dramatic index of refraction change occurring around a thermotropic liquid crystalline phase transition, we demonstrate a 40-fold signal-to-noise ratio enhancement for gold nanoparticles imaged in 4-Cyano-4'-pentylbiphenyl (5CB) liquid crystals over those in a water environment. We studied the photothermal signal as a function of probe laser polarization, heating power, and sample temperature quantifying the optimal enhancement. This study established photothermal microscopy as a valuable technique for inducing and/or detecting local phase transitions at the nanometer scales.

With growing research in nanoparticles and their uses in nanosystems such as biological sensing and therapeutics[1], there is a need for developing simple, versatile methods to characterize them at the single-nano-object level. The most common techniques are based on luminescence but are often subject to photobleaching or blinking. Alternative, more stable techniques have been developed relying solely on absorption of nano-objects[2]. One such highly sensitive technique is photothermal heterodyne imaging (PHI) [3], which has been shown to image Au nanoparticles as small as 1.4 nm, semiconductor nanocrystals[4] or single molecules of Black-Hole-Quencher-DNA construct[5]. To further improve the sensitivity of PHI[6], it is important to choose a medium that has the greatest refractive index variations with temperature $\partial_T n$ [7, 8]. In this context, a medium displaying sudden refractive index variations around a phase transition should offer high signal sensitivities.

Moreover, PHI enables probing of temperature dependent effects at sub-wavelength length scales around heated nanoparticles. Indeed, many new materials with advantageous physical properties are produced by incorporating nanostructures. PHI can thus offer unique insights into how the material surrounding a nanostructure behaves at the nanoscale[9], which becomes very interesting when the material undergoes a phase transition.

Herein, we show that thermotropic phase transitions in 4-Cyano-4'-pentylbiphenyl (5CB) liquid crystals can provide a 40-fold enhancement of the photothermal signals. Moreover, using the temperature dependence of the signal enhancement measured on individual gold nanoparticles we probe the nematic-to-isotropic phase transition occurring at the nanoscale level.

PHI microscopy uses a tightly focused time-modulated heating beam (532 nm) superimposed with a non-resonant probe beam (633 nm)[3, 8]. An absorbing nano-object in the focal volume produces a time-modulated refractive index profile of amplitude $\Delta n(r)$ with $r$ the distance to the nanoparticle center. This profile is given by $\Delta n(r) = \Delta T(r) \partial_T n(r)$ with $\Delta T(r) = \Delta T_S f(r)$ the temperature profile around the particle, $\Delta T_S$ the temperature rise at the surface of the particle (proportional to the absorbed power $P_{abs}$), and $f(r)$ a function which depends on the heat diffusion properties in the medium. The interaction of the probe beam with the refraction index profile produces a scattered field with sidebands at the modulation frequency. A lock-in detection system is used to detect the beatnote of the forward scattered field with the transmitted probe field at the modulation frequency.

5CB is a well-characterized liquid crystal, in both thermal and optical properties as well as phase behavior with advantageous photothermal properties[10-11]. As shown in Figure 1a, its index of refraction is highly dependent on both temperature and polarization with respect to the nematic axis. Indeed in the isotropic phase $\partial_T n_{iso} \sim -6 \cdot 10^{-4}$ K$^{-1}$ a value significantly larger than that of water ($\sim -10^{-4}$ K$^{-1}$) and comparable to that of viscous silicone oils (between $-3.5 \cdot 10^{-4}$ K$^{-1}$ to $-5.0 \cdot 10^{-4}$ K$^{-1}$). Well below the phase transition temperature ($T_C \sim 32°C$, see methods)[12], a linearly polarized beam parallel or orthogonal to the nematic axis will experience distinct indexes of refraction $n_{//}$ and $n_\perp$ respectively. Interestingly, $\partial_T n_{//}$ is four times larger than in the isotropic phase while $\partial_T n_\perp$ is weaker and have opposite sign. Around $T_C$ the index of refraction displays sharp variations with $T$ where large PHI signal enhancements are expected. Figure 1 d-e shows PHI images of 28 nm diameter gold beads spin coated on a glass coverslip and covered by an aligned 7-10 μm thick layer of 5CB (see methods). The PHI signal $S_{PHI}$ represents the magnitude of demodulated signal by the lock-in amplifier. Images are recorded with a probe beam polarized along the nematic axis at sample temperature $T_{sample}$ of 23°C (Figure 1d) and 31°C (Figure 1e). The heating intensity was 10 kW/cm² corresponding to an absorbed power of $P_{abs} \sim 100$ nW for an average size bead and an average temperature rise at the surface of the particle[13] of $\Delta T_S \sim 2.5$ °C. This implies that, in the case of Figure 1d the maximum temperature at the vicinity of the particles $T_{sample} + \Delta T_S$ does not reach $T_C$, while in Figure 1e the temperature is modulated around $T_C$ such that the liquid crystal experiences a local phase transition in the vicinity of the heated beads. In order to compare the signal enhancement due to the use of the liquid crystal we recorded images of the same size particles covered by silicone oil instead of 5CB. For quantitative comparison, the images are recorded at the same $\Delta T_S$, i.e. the same absorbed power $P_{abs}$ (Figure 1c) and not at the same heating intensity, since the gold nanoparticle absorption cross-section depends on the medium index of refraction[14]. For this purpose, we record direct absorption images of the heating beam by the nanoparticles in the different media and adjust the heating intensity to achieve the desired $P_{abs}$. A clear signal enhancement is observed with liquid crystals. In the case of Figure 1d (well below phase transition), the enhancement is due to the fact that 5CB has $\partial_T n$ four times larger than silicone oil. Figure 1e reveals that around the transition an additional enhancement is obtained. The signal enhancement is more striking if the signals obtained with 5CB are compared to that obtained in water. At the same $P_{abs}$ the nanoparticles cannot be detected in water and one needs to increase the heating intensity by four fold to barely image them (Figure 1b). We obtained that the observed enhancement between water medium and 5CB reaches 40-folds in average.

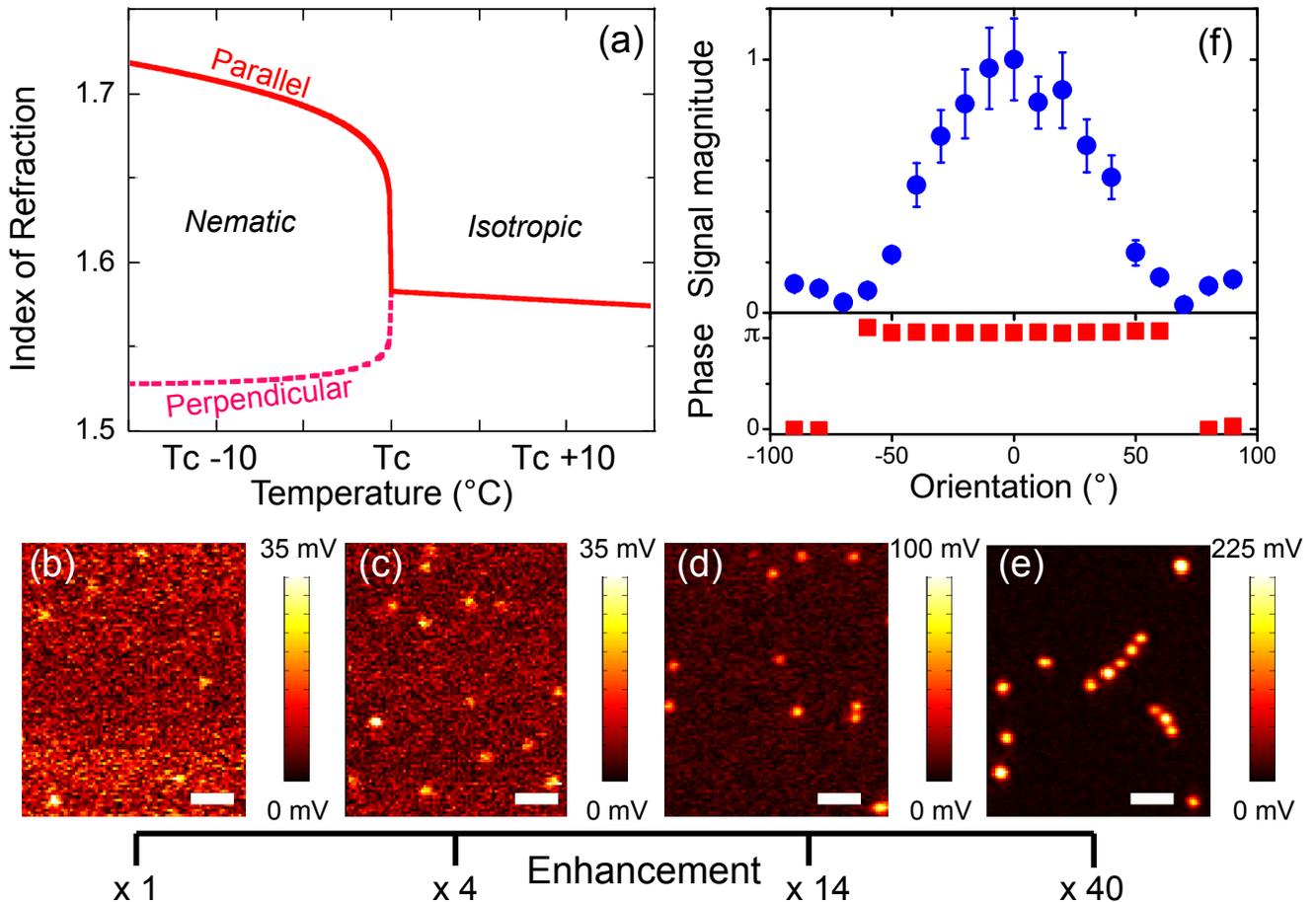

Figure 1: (a) Index of refractions (parallel and perpendicular to the nematic axis) of 5CB as function of temperature. Above the phase transition temperature $T_C$ the index of refraction is isotropic. (b-e) Photothermal images of 28 nm gold nanoparticle recorded in different media : (b) in water, (c) in silicone oil, (d) in 5CB at 23°C (e) in 5CB at 31°C. The integration time is 12 ms and scale bar 1 μm. The heating intensity was adjusted in silicone oil and 5CB to induce an absorption power leading to 2.5°C temperature rise of at the surface of the nanoparticle. In water, the absorbed power is 3 times larger. (f) Photothermal normalized signal magnitude and phase as a function probe beam polarization with respect to the nematic axis.

Figure 1f shows the signal magnitude dependence on the probe beam polarization orientation of nanoparticles imaged in 5CB medium. As expected, the maximum signal is obtained along the nematic axis (the reference orientation), drops to the noise level at ±70° and then increases to a lower maximum for orthogonal orientations. Furthermore, the phase of the demodulated signal delivered by the lock-in amplifier displays a π shift at ±70° (Figure 1f). These observations are a consequence of the opposite variations of $n_{//}$ and $n_\perp$ with temperature (Figure 1a). The detailed modelization of the signal is out of the scope of this letter and will be published elsewhere.

In order to unambiguously state that the local phase transition contributes to the observed signal enhancement, we now study the signal dependence with $\Delta T_S$ (i.e. with $P_{abs}$). Within the simplified model of a plane wave at the focus of the beams and of small index of refraction variations, the PHI signal can be approximated to $S_{PHI} \propto \int d^3\vec{r} \Delta n(r)$. For single-phase mediums (silicone oils, water), the index of refraction gradient can be considered constant for reasonable temperature excursions. In this case, $S_{PHI} \propto \Delta T_S \partial_T n \int d^3\vec{r} f(r)$ scales linearly with $\Delta T_S$ as can be seen on Figure 2a (open symbols). In a phase changing medium, the former proportionality does not hold (see Figure 2 filled symbols) because $\partial_T n$ is function of the temperature thus function of the distance to the particle. Indeed, at sample temperature well below $T_C$, the signal first increases linearly with $\Delta T_S$ at a steeper assent than in silicone oil due to the greater $\partial_T n$ of the 5CB nematic phase. Then, a super-linear growth in the signal is clearly visible owing to the sharp increase of $\partial_T n_{//}$ induced by the phase transition that occurs in the environment of the nanoparticle. As the power is further increased, the relative proportion of material undergoing a phase transition in the probe beam focal volume increases giving more of a signal enhancement. The latest starts to level off when the heating power induces a phase transition in the whole volume defined by the probe beam. One can estimate the local phase transition temperature (~32°C) from $\Delta T_S$ at which the signal starts to deviate from linearity. For larger absorbed powers, the signal increases again linearly but with a lower slope given by $\partial_T n_{iso}$, of the isotropic phase. This overall behavior is clearly understood from Figure 2b-c where we have solved the heat equation diffusion in a medium experiencing a phase transition[15] using the thermal parameters of 5CB[10] and an averaged heat conductivity for the nematic phase in order to plot $\Delta T(r)$ (Figure 2b) and $-\Delta n_{//}(r)$ (Figure 2c) for different $\Delta T_S$. One can see that it is the rapid drop of $\Delta T(r)$ with $r$ which limits the extent of the region undergoing a phase transition around the nanoparticle and imposes the signal evolution with $\Delta T_S$.

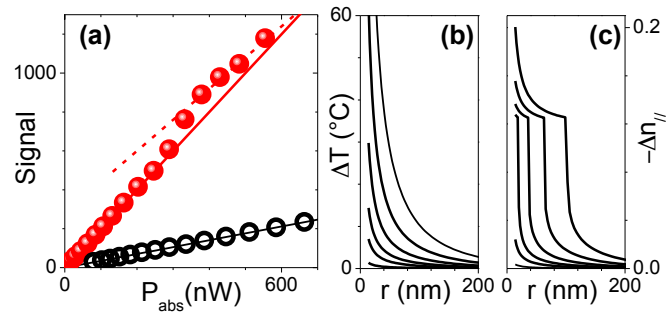

Figure 2: (a) Signal evolution of individual gold nanoparticles recorded in 5CB (red filled symbols) and in silicon oil (black open symbols) as a function of $P_{abs}$ for $T_{sample} = 23°C$. (b-c) Theoritical temperature rise (b) and index of refraction variations parallel to the nematic axis $-\Delta n_{//}(r)$ as a function of the distance to the particle center for $P_{abs} = 20, 100, 200, 400, 800$ and $1600$ nW.

Noteworthy, the contribution of the phase change to the overall PHI signal shown above is not fully exploited in the experimental configuration of Figure 2 ($T_{sample}$ well below $T_C$). Indeed, if all the material within the probe beam can transitioned without the need of high $P_{abs}$, a greater signal enhancement should be obtained. This can be accomplished by using small $\Delta T_S$ and increasing the overall temperature of the sample near $T_C$. In Figure 3 we recorded $S_{PHI}$ as a function of $T_{sample}$ (22°C to 34°C) while maintaining $\Delta T_S$ to a fixed small value (0.5°C). As expected from the sharp variation of $\partial_T n$ the enhancement of the signal when $T_{sample}$ is raised from 23°C to 31°C is larger using $\Delta T_S = 0.5$ °C ($\approx 4.3$ Figure 3) than $\Delta T_S = 2.5$ °C ($\approx 2.8$ Figure 1d-e).

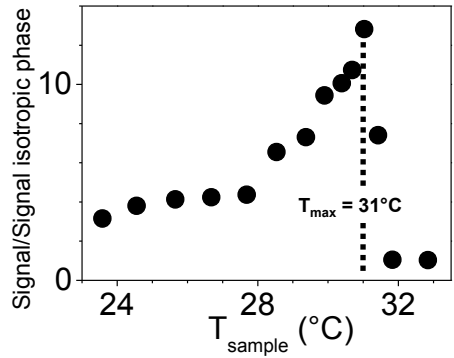

Figure 3: PHI signal measured in 5CB as a function of increasing global sample temperature for 28 nm NPs and normalized by the signal obtained in the isotropic phase. The heating intensity is adjusted to induce a 0.5°C temperature rise of at the surface of the nanoparticle. The maximum signal enhancement is obtained close to the phase transition temperature.

At low $T_{sample}$, $S_{PHI}$ increases with increasing temperature accordingly to the variations of $\partial_T n$ with temperature. Then, as $T_{sample}$ reaches $T_C - \Delta T_S$ the local phase transition, which induces the sharp increase in $\partial_T n$, starts to occur at the surface of the nanoparticle and leads to a pronounced signal enhancement. The maximum signal is obtained at the temperature $T_{MAX}$ such that $\Delta T_S$ induces a phase transition in the whole probe beam volume (i.e. when $T_C$ is reached at the boundaries of the volume).

Approximating the temperature profile around the nanoparticle by $T_{sample} + \Delta T_S \frac{a}{r}$, one should thus find

$T_{MAX} = T_C - \Delta T_S \frac{a}{R_{beam}}$ with $a$ the radius of the nanoparticle and $R_{beam}$ the characteristic size of the confocal probed volume. Since $a \ll R_{beam}$, $T_{MAX}$ should be a direct measurement of the local phase transition temperature $T_C$. Experimentally, one finds $T_{MAX} = T_C = 31°C \pm 0.5°C$ in agreement with the global phase transition temperature of 5CB ($32°C \pm 1°C$)[12] measured on our sample (7-12 µm thick) by a birefringence technique.

In conclusion, we have shown that the sensitivity of photothermal microscopy can be enhanced by up to 40-fold near the phase transition of 5CB liquid crystals. We demonstrate that the fraction of material phase transitioning within the confocal volume determines the signal enhancement. Optimal enhancement is obtained for probe beam polarization along the nematic axis and sample temperature close to the phase transition of the liquid crystals. Further experiments aim to the detection of tiny nano-absorbers and to use nano-heating to study order dynamics in such media.

EXPERIMENTAL METHODS

The PHI setup used here corresponds to the forward direction scheme[3, 8]. A non-resonant probe beam (HeNe, 632.8 n) and an absorbed heating beam (532 nm, frequency doubled Nd:YAG laser) are overlaid and focused on the sample by means of a high NA microscope objective[16] (60×, NA=1.49, oil immersion). The intensity of the heating beam is modulated at a frequency Ω (a few 100 kHz) by an acousto-optic modulator. The interfering probe-transmitted and forward-scattered fields are efficiently collected using a second microscope objective (60×, water immersion) on a fast photodiode and fed into a lock-in amplifier in order to extract the beatnote signal at Ω. Photothermal images are obtained by raster scanning of the samples by means of a piezoscanner stage. The probe beam power (~100-200µW) induces a particle surface heating (< 2°C) which is not modulated and does not contribute to the PHI signal. A resistor heating element with 11 mm optical clearance and two objective heaters were used to control the temperature within 0.1°C (Bioscience Tools).

The liquid crystal 4-Cyano-4'-pentylbiphenyl (5CB) was purchased from Hebei Maison Chemical Co., LTD. Gold nanoparticles (*Nanopartz TM Inc.*) of 27.8±1.5 nm diameter as determined by TEM were diluted with (2%) aqueous dispersions of polyvinyl alcohol (PVA) and spin-coating on a plasma cleaned coverslip. A 1% PVA solution was spin-coated onto a second plasma- cleaned coverslip before being lightly scratched with velvet to induce large, highly aligned domains in the 5CB liquid crystal that is placed immediately after (1.5 µL drop). The PVA coated coverslip that contained the nanoparticles was then turned upsidedown (coated side facing down) and placed on top sandwiching the 5CB between both coverslips. Once the 5CB fills the space between the coverslips, the thickness is estimated to be between 7 and 10 µm, epoxy was then used to seal and fix the coverslips in place. As a control, samples were prepared by the same technique (without epoxy) with viscous silicone oil or water.

ACKNOWLEDGMENT. We warmly thank Cécile Leduc and Jonah Shaver for helpful discussions. This work was funded by the Agence Nationale de la Recherche, Région Aquitaine, and the European Research Council.